\documentclass[sigconf]{acmart}
\usepackage[utf8]{inputenc}

\usepackage{soul}
\usepackage{xcolor}
\usepackage{xspace}
\usepackage{url}
\usepackage{graphicx}
\usepackage{listings,multicol}
\usepackage[caption=false]{subfig}  
\usepackage[export]{adjustbox}
\usepackage{amsmath}

\usepackage{amssymb}
\usepackage{textcomp}
\usepackage{breakurl} 
\usepackage{listings}
\usepackage{hyperref}
\usepackage{cleveref}
\usepackage{algorithm}
\usepackage[noend]{algpseudocode}
\usepackage{dblfloatfix} 
\usepackage{enumitem}
\usepackage{kantlipsum}
\usepackage{multirow}

\usepackage{color}
\newcommand*{\colordefine}[2]{%
  \expandafter\colordefineAux#2{#1}%
}
\newcommand*{\colordefineAux}[3]{%
  \definecolor{#3}{#1}{#2}%
}

\definecolor{annotationsColor}{HTML}{808002} 
\definecolor{classColor}{HTML}{000000} 

\usepackage{booktabs}
\usepackage{diagbox}
\usepackage{fixfoot}

\makeatletter
\def\lst@makecaption{%
  \def\@captype{table}%
  \@makecaption
}
\makeatother

\newcommand{\eg}{\emph{e.g.,}\xspace}

\newcommand{\toolname}{\emph{Reflekt}\xspace}
\newcommand{\toolnameAPI}{\texttt{Reflekt}\xspace}
\newcommand{\smarttoolname}{\texttt{Smart\-Reflekt}\xspace}
\newcommand{\kotless}{\emph{Kotless}\xspace}

\makeatletter
\newcommand{\linebreakand}{%
  \end{@IEEEauthorhalign}
  \hfill\mbox{}\par
  \mbox{}\hfill\begin{@IEEEauthorhalign}
}
\makeatother

\acmConference[ICSE 2022]{The 44th International Conference on Software Engineering}{May 21–29, 2022}{Pittsburgh, PA, USA}

\title{\toolname: a Library for Compile-Time Reflection in Kotlin}
\author{Anastasiia Birillo}
\affiliation{
  \institution{JetBrains Research}
  \city{Saint Petersburg}
    \country{Russia}
}
\email{anastasia.birillo@jetbrains.com}

\author{Elena Lyulina}
\affiliation{
  \institution{JetBrains Research}
  \city{Saint Petersburg}
    \country{Russia}
}
\email{elena.lyulina@jetbrains.com}

\author{Maria Malysheva}
\affiliation{
  \institution{JetBrains Research}
  \institution{Saint Petersburg State University}
  \city{Saint Petersburg}
    \country{Russia}
}
\email{maria.malysheva@jetbrains.com}

\author{Vladislav Tankov}
\affiliation{
  \institution{HSE University}
  \institution{JetBrains Research}
  \city{Saint Petersburg}
    \country{Russia}
}
\email{vladislav.tankov@jetbrains.com}

\author{Timofey Bryksin}
\affiliation{
  \institution{HSE University}
  \institution{JetBrains Research}
  \city{Saint Petersburg}
    \country{Russia}
}
\email{timofey.bryksin@jetbrains.com}

\copyrightyear{2022}
\acmYear{2022}
\setcopyright{acmlicensed}\acmConference[ICSE-SEIP '22]{44nd International Conference on Software Engineering: Software Engineering in Practice}{May 21--29, 2022}{Pittsburgh, PA, USA}
\acmBooktitle{44nd International Conference on Software Engineering: Software Engineering in Practice (ICSE-SEIP '22), May 21--29, 2022, Pittsburgh, PA, USA}
\acmPrice{15.00}
\acmDOI{10.1145/3510457.3513053}
\acmISBN{978-1-4503-9226-6/22/05}

\begin{document}

\begin{abstract}
Reflection in Kotlin is a powerful mechanism to introspect program behavior during its execution at run-time.
However, among the variety of practical tasks involving reflection, there are scenarios when the poor performance of run-time approaches becomes a significant disadvantage. 
This problem manifests itself in \kotless, a popular framework for developing serverless applications, because the faster the applications launch, the less their cloud infrastructure costs.
In this paper, we present \emph{Reflekt} --- a compile-time reflection library which allows to perform the search among classes, object expressions (which in Kotlin are implemented as singleton classes), and functions in Kotlin code based on the given search query. It comes with a convenient DSL and better performance comparing to the existing run-time reflection approaches.
Our experiments show that replacing run-time reflection calls with \emph{Reflekt} in serverless applications created with \kotless resulted in a significant performance boost in start-up time of these applications.
\end{abstract}

\maketitle

\section{Introduction}\label{sec:introduction}

\kotless~\cite{tankov2021infrastructure} is a framework created within JetBrains to facilitate the development of serverless applications in Kotlin.
Serverless computing is a concept of developing and deploying cloud applications as a set of stand-alone functions, which are executed on demand and automatically scaled if needed~\cite{castro2017serverless}. 
Such an approach suits a wide range of applications, and compared to developing them as traditional server-side applications it enables better resource management while being scalable, reliable, and cost-effective~\cite{castro2017serverless}. 
\kotless aims to simplify the process of developing and deploying such serverless applications.
The framework focuses on reducing the routine of serverless deployment by automatically generating deployment code straight from the source code of the application itself.
\kotless is fully open-source~\cite{kotlesslink} and has an active and rapidly growing community.

Being a Kotlin-specific framework, \kotless is also written in Kotlin, an open-source statically typed programming language that targets JVM, JavaScript, and native platforms via LLVM~\cite{llvm}.
Kotlin is fully interoperable with Java, enabling to reuse all Java features and existing libraries.

Implementation-wise \kotless heavily relies on the reflection mechanism.
In general, reflection can be defined as the ability of a program to manipulate as data something representing the state of the program during its own execution~\cite{bobrow1993clos}, and it is a common functionality in many programming languages, including but not limited to  
Python, Ruby, Go, R, Java, and, therefore, Kotlin.
While reflection provides developers with a variety of possibilities, including the ability to manipulate with the program structure, one of its useful features is introspection --- the ability of a program to examine its own state and structure~\cite{maes1988meta}. 
One of the examples of program introspection is type identification, which is implemented via the \texttt{instanceof} operator in Java and the \texttt{is} operator in Kotlin, respectively.

However, such a powerful mechanism as reflection in Java, along with all the possibilities, brings its own drawbacks and is usually advised to be used carefully~\cite{li2019understanding}.
Among them are security restrictions, coming from the opportunity to load and execute classes at run-time. 
Some virtual machines, \eg used in Android development, do not provide a secure environment for code supplied dynamically~\cite{zhauniarovich2015stadyna}, and Google is strongly opposed to using this feature in Android applications~\cite{androidsecurity}.
Reflection also hampers the usage of GraalVM~\cite{GraalVM}, a virtual machine that supports ahead-of-time compilation~\cite{GraalVMAOT} for faster program start-up and lower memory consumption~\cite{GraalVMNativeImage}.
Also, there are challenges with static analysis of the code, since it is fundamentally hard to predict the behavior of the code that uses reflection, which can be done only under significant assumptions~\cite{landman2017challenges}.
Last but not least is performance degradation. 
That is caused by the fact that reflection-based operations require types that must be dynamically resolved, and to do that, JVM should load all the necessary classes at run-time, slowing down the application.
Poor performance of Java reflection has been widely studied~\cite{landman2017challenges, tudose2013java, li2019understanding} and this is one of the main reasons developers try to use reflection only when it is strictly necessary.

In \kotless, reflection is used to handle the event-driven way serverless applications are written in. 
In this case, it is limited to the task of collecting all the classes, objects (Kotlin classes that can have only one instance), and functions that satisfy some search condition across all the application's source code. 
Classes that implement a specific interface or functions with a specific list of arguments and return types might serve as an example of such conditions.

The default way of doing this in Kotlin is 
Java Reflection API and libraries such as \textit{reflections}~\cite{reflections}, which provide a convenient domain-specific language.
This approach allows collecting all required entities at run-time by scanning the whole classpath and checking the classes against the given search constraint.
However, as mentioned above, iterating over all the accessible entities during the program run might be very inefficient~\cite{landman2017challenges, tudose2013java, li2019understanding}. 
Compile-time approaches, such as Java Annotation Processor~\cite{annotationProcessing}, are another possible option. 
Such tools search for the required entities only once during the compilation stage, and the result of this search can be stored somewhere to make the fetching of the required entities at run-time much faster.
However, to make such a compile-time search possible, all the entities in the code should be marked in advance with specific annotations, which often makes this approach rather inconvenient to use. 
The tedious work of annotating entities along with the code becoming cluttered with annotations prevented \kotless developers from using this approach in the first place.

Therefore, there is a need for a fast and at the same convenient way to employ reflection methods.
The former can be achieved by moving the process to compile-time, and the latter may be solved by developing a user-friendly DSL that requires minimum efforts to switch from a run-time library, such as \textit{reflections}, to the new, compile-time approach.
Unlike Java, in Kotlin all the above is possible thanks to the capability to write compiler plugins. 

In this paper, we present a Kotlin reflection library called \toolname~\cite{reflekt} that overcomes the flaws of the standard Java reflection approach and provides the means to find classes, object expressions, and functions by a given condition without degrading the application's performance.
Instead of relying on the JVM reflection infrastructure, \toolname performs compile-time resolution of reflection queries using the Kotlin compiler analysis infrastructure, providing a convenient reflection API without actually using reflection.
The search condition might be specific supertypes, annotations, and signatures of code elements, which is a typical use case for \kotless, as well as a custom user condition.

We evaluated \toolname on two applications that utilize \kotless and achieved significant acceleration in their start-up.
For the first project, replacing the run-time reflection approach with \toolname decreases its start time by 13.8\%, and for the second one, the results are even better with the overall improvement of 17\%.
At the same time, the compile-time approach of \toolname did not affect the compilation time very much, only increasing it by about 1\% for both applications. 

The rest of this paper is organized as follows. 
\Cref{sec:background} describes existing approaches to the problem of searching for entities in the source code, motivates the development of \toolname, and briefly introduces the Kotlin compilation process. 
\Cref{sec:design} presents the pipeline of \toolname and its implementation details, while in  \Cref{sec:evaluation} we discuss the practical evaluation of \toolname.
Finally, \Cref{sec:conclusion} sums up the work and discloses our future plans.

\section{Background}\label{sec:background}

In this section, we provide a motivational example of a reflection-based \kotless feature and explore several ways it can be implemented with the existing run-time and compile-time reflection approaches. We also describe the approach to finding classes and functions in the code proposed in this paper and highlight existing real-world projects that could benefit from such an approach.

\subsection{Motivating Example}\label{subsec:motivating-example}

\kotless~\cite{tankov2021infrastructure} is a framework for developing and deploying serverless applications to Amazon Web Services (AWS) and Microsoft Azure clouds.
From the user's point of view, \kotless is almost invisible --- it is integrated as a Gradle plugin and generates all the necessary deployment code automatically. 
To do that, \kotless introduces a DSL based on annotations and marker interfaces, that allow users to define how their application should be deployed to the cloud~\cite{tankov2019kotless}.
The performance of \kotless fully depends on the ability to find such annotations and marker interfaces in the users' code, and this task is a perfect example of what reflection does. 
At the same time, serverless applications should start and work as fast as possible since they could be launched hundreds of times a day, and the longer they work, the more their end-users wait, the more CPU time is consumed, and the more money is eventually spent on cloud infrastructure~\cite{adzic2017serverless}.
Hence, in such a case, using run-time reflection, which is known for its poor performance~\cite{tudose2013java, landman2017challenges}, can be quite costly.

Let us look closer at a concrete example of how reflection is used in \kotless. 
One of the features of \kotless is the ability to define events that will be repeatedly executed after a given time interval. 
To do this, a developer should mark the desired function with the \texttt{@Scheduled} annotation: 

\begin{figure}[!h]
    \includegraphics[width=0.62\columnwidth, left]{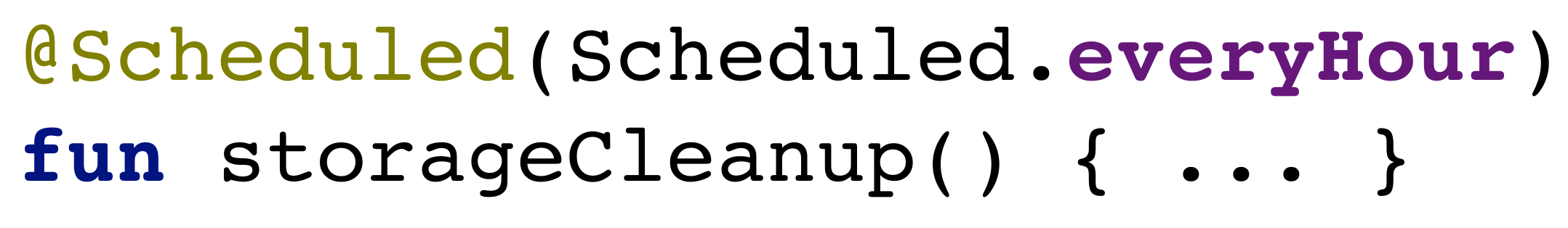}
\end{figure}

The job of \kotless is to find such annotations in the application code and generate appropriate cloud infrastructure code to make this repeated execution possible.
Now, having this specific example in mind, let us discuss in detail how this task of finding entities in code could be implemented in Kotlin. 

\subsection{Existing Reflection Approaches}
\subsubsection{Run-time reflection}
The built-in approach to solving this task is Java Reflection API~\cite{javaReflectionApi}, which consists of two main parts: objects that represent various parts of the program and tools for extracting these objects.
To fetch information about a class, Java introduces the \texttt{Class} class with a set of methods to introspect it, \eg to find a required method or check for specific annotations.
Class loaders scan the classpath for a needed \texttt{Class} and load it dynamically on demand. 
Therefore, the task of finding classes, objects, and methods by specific conditions is already entirely covered by the Java reflection infrastructure.
Employing it for our motivating example, the solution will result in the following code fragment:

\begin{figure}[H]
    \includegraphics[width=0.95\columnwidth, left]{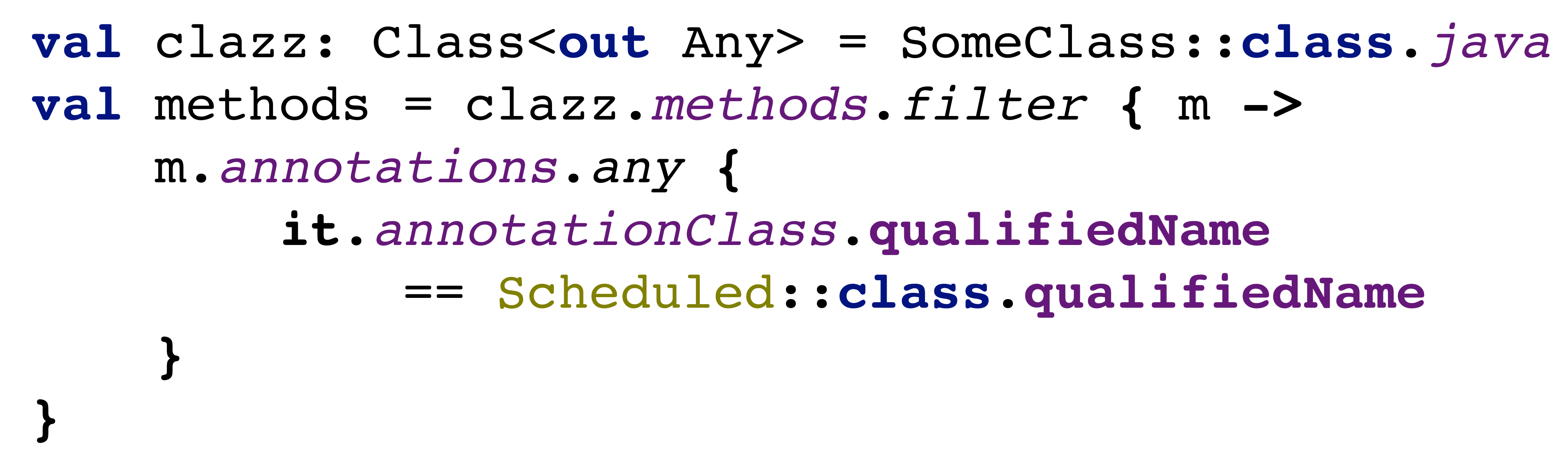}
\end{figure}

For each \texttt{Class} from the classpath we should extract the set of its methods' annotations and check whether \texttt{@Scheduled} is among them.

Several libraries utilize Java Reflection API or employ a similar approach to achieve this task at run-time as well. 
For example, the aforementioned \textit{reflections}~\cite{reflections} is a popular open-source library, 
which comes with a clear and intuitive DSL.
Getting back to the motivating example, the task of finding all the \texttt{@Scheduled} functions is currently solved in \kotless with the \textit{reflections} library like this: 

\begin{figure}[!h]
    \includegraphics[width=0.95\columnwidth, left]{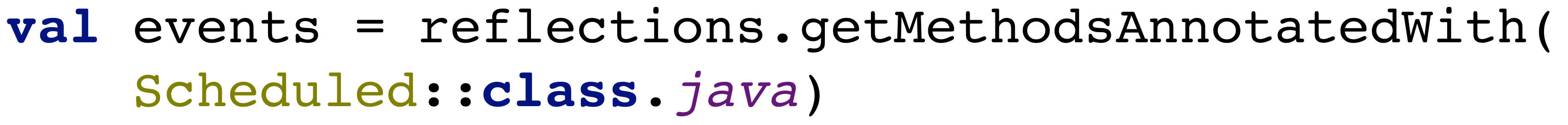}
\end{figure}

Nevertheless, such a convenient tool comes with several disadvantages inherent to all approaches of this type, with the poor performance being the most notable.
This is caused by the necessity to scan the whole classpath at run-time every time the program runs, which results in significant performance overhead. 
In fact, the inefficiency of Java Reflection API is a well known problem~\cite{tudose2013java}. 
Historically, one of the main reasons this API was added in Java was the ability to introspect programs specifically at run-time.
But in certain tasks, such as ours, where run-time execution is not a must, an alternative might be more practical.

\subsubsection{Compile-time approaches}\label{subsubsection:compile-time}
Another option would be to use approaches that search for the entities at compile-time and store the information about the entities of interest to use it later at run-time. 
Here, the scan of the classpath is performed at compile-time, not slowing down the application runs.

One of the examples of such an approach is Java Annotation Processing~\cite{annotationProcessing} and several libraries~\cite{classindex, scannotation, classgraph}, which are based on it or follow the same strategy. 
When the compiler encounters a specific annotation, it runs an annotation processor created to handle this specific annotation.
The annotation processor can then handle the annotated entity and get the same information Java Reflection API provides out of it. 
This information is stored for later use at run-time.  

Therefore, for this approach to work, one needs to (1) mark all the desired entities in code with an appropriate annotation (\eg if a developer wants to find all classes with a given supertype, this should be expressed with an annotation), and (2) for each annotation, implement a custom annotation processor with the desired behavior and register it in the compiler.
Usually, this results in a lot of extra work and makes the code cluttered with endless annotations. 

The task of finding all the \texttt{@Scheduled} functions from the motivating example could be solved using Java Annotation Processing as well.
For that, a subclass of \texttt{AbstractProcessor} needs to be created with a custom implementation of the \texttt{process()} method.
It defines which annotations to search, which conditions to satisfy, and how to handle the entity afterwards:

\begin{figure}[!h]
 \includegraphics[width=1\columnwidth]{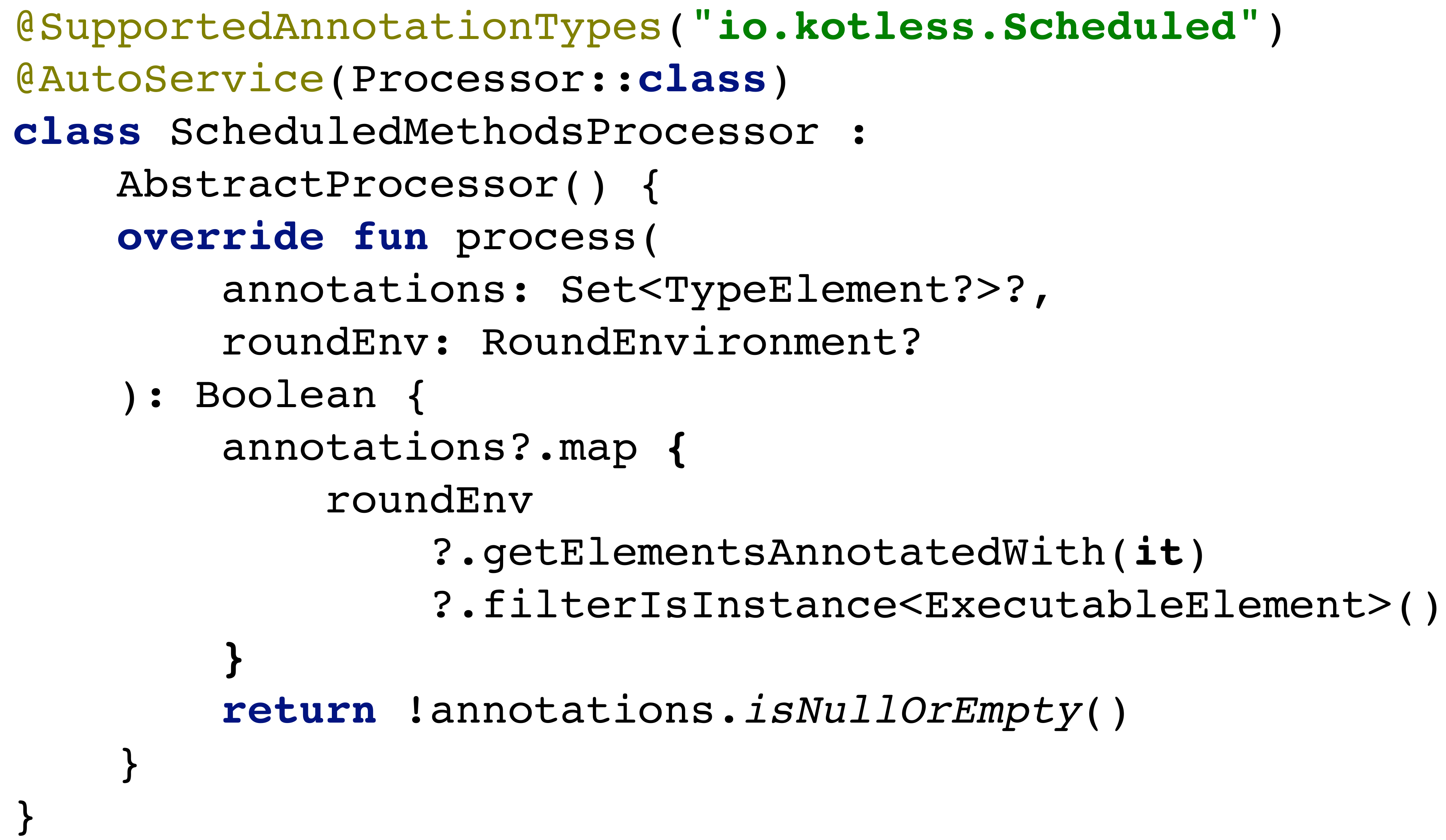}
\end{figure}

In addition to the annotation processing technique, there are several libraries that optimize the time-consuming classpath scanning inherent to run-time approaches. 
For example, the \textit{ClassGraph} library~\cite{classgraph} allows to scan the classpath searching for classes that match some criteria, be it a specific annotation or a supertype. 
The idea here is just to remember the found classes at compile-time, thus reducing the size of the classpath needed to be scanned at run-time with Java reflection.
Nevertheless, the need to implement serialization and deserialization of results manually for every search significantly increases the amount of code the users have to write.

Therefore, the existing compile-time approaches, while being much faster than run-time ones, cannot offer a way to solve the given task in just a few lines of code. 
They either require putting additional annotations everywhere in the source code and implementing annotation processors for every concrete annotation, or spending time on storing and loading the scan results in order to make them visible at run-time. Besides, the existing libraries often do not cover the more general task of finding classes, functions, and objects satisfying a custom search condition. 

\begin{table*}[h]
    \begin{center}
        \begin{tabular}{p{8.5cm} p{8.5cm}}
            \toprule
               \textbf{Description}   & \textbf{Reflection context}  \\ \midrule 
               \textbf{Kotlin Faker}~\cite{kotlinFaker} is a tool for generating realistically looking fake data such as names, addresses, banking details, and many more, that can be used for testing purposes.
               The tool performs ahead-of-time compilation to a GraalVM~\cite{GraalVM} native image that speeds up the start-up and reduces the memory usage.
& 

Kotlin Faker uses Java Reflection API to find and access classes and functions properties, as well as to collect code entities by specific annotations.
However, the only way to make reflection work with GraalVM is to manually create a configuration file, listing a pre-written description of the entities that are needed to be found~\cite{GraalVMReflection}.  
\\ \midrule

\textbf{Rawky}~\cite{rawky} is an editor for the pixel art graphics~\cite{silber2015pixel}.
It has a graphical user interface (GUI) to interact with the image, \eg repeat or cancel actions, add layers to the picture, etc. & 

Run-time reflection in this project is used to initialize the editor's GUI, \eg getting the options for the windows and panels. 
Here, it is done using the \textit{reflections}~\cite{reflections} library.
\\ \midrule

\textbf{Detekt}~\cite{detekt} is a static code analysis tool for Kotlin.
It finds various code smells~\cite{fowler2018refactoring}, code style violations, calculates code quality measures, \eg code complexity.
To perform such analyses, Detekt operates with the syntax tree, provided by the Kotlin compiler. &

Detekt implements its analyzers as sets of rules. Each of them is represented by a class implementing a specific interface. Reflection is used in this project during testing (via the \textit{reflections} library) to collect all such classes and validate the analyzers.
 \\ \midrule

\textbf{Kotest}~\cite{kotest} is a flexible testing tool for Kotlin with the multiplatform support (JVM and JavaScript). 
Kotest consists of several different parts: a framework that allows to define and execute tests, an assertion library with a diverse set of over 300 assertions, and a module for property-based testing~\cite{fink1997property}.& 

Kotest is supposed to be used as a library in projects to define and run tests. 
It employs the ClassGraph~\cite{classgraph} reflection library to set up custom user configurations, as well as to search and apply additional automatic configurations to the project.
\\ \midrule

\textbf{Ktor}~\cite{ktor} is an asynchronous flexible framework for creating microservices and web applications. 
The framework provides a special DSL to create web applications of different scale and complexity. &

Ktor does not use reflection to avoid the corresponding slowdown, however, it can also benefit from using \toolname. 
The framework provides an HTTP API, which is based on Spring~\cite{springBoot} annotations.
In large projects, it could be quite inconvenient to define hundreds of such functions and call them one by one from the main application context. 
With \toolname, it is possible to find all Spring annotations and construct Ktor functions automatically at compile-time.
              \\\bottomrule
            \end{tabular}
            \vspace{0.2cm}
            \caption{Several existing projects from different domains that 
            can benefit from using \toolname.}
        \label{tab:examples}
    \end{center}
\end{table*}

\subsubsection{The \toolname approach}

Run-time and compile-time \linebreak approaches complement each other, with the former providing a convenient DSL but being slow, and the latter being fast but requiring the user to write a lot of extra code. 
Fortunately, there is a way to combine the advantages of both approaches.

Unlike Java, Kotlin provides a way to write plugins for its compiler, thus extending its functionality.
It is possible to intervene into the compilation process to scan the compiling code for the required entities, filter them to match the search criteria, and collect them.
Thus, such an approach preserves the efficiency of existing compile-time methods.
Also, there is no need to mark the code with annotations, since we can use all the information available to the compiler. 
Moreover, it is possible to relieve the users of the obligation to store search results between the application compilation and running.
The Kotlin compiler allows us to modify the compiled code, therefore, all the found entities can be directly written into the code's intermediate representation, without the users even noticing it.
All of the above allows us to implement a Kotlin compiler plugin with a concise and convenient DSL to search for classes, objects, and functions satisfying custom search conditions just in a few lines of code similar to run-time approaches. 

Returning to our motivating example and the task of finding all the functions with the \texttt{@Scheduled} annotation, with the \toolname approach it could be solved like this:

\begin{figure}[!h]
    \includegraphics[width=0.83\columnwidth]{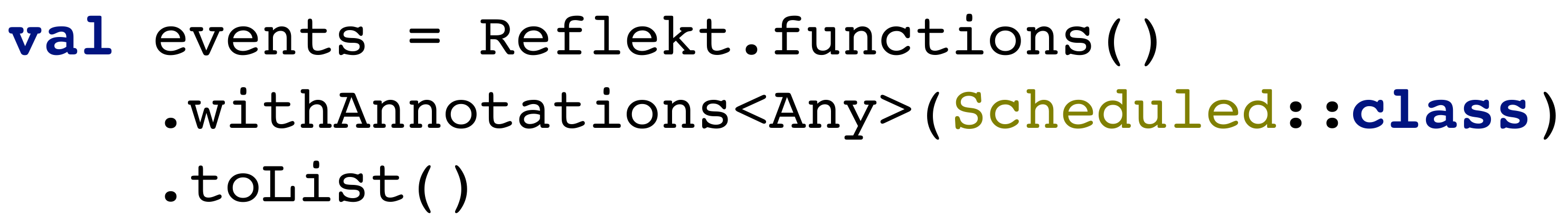}
\end{figure}

\subsection{Use Cases of \toolname}
Although the original motivation behind the development of \toolname was to overcome the performance slowdown caused by run-time reflection in \kotless, \toolname can be useful outside of \kotless as well.
In a recent study, Landman et al.~\cite{landman2017challenges} reported that among the 461 open-source projects they looked at, more than 96\% of them contained Java Reflection API calls. 
Almost 70\% of these projects used reflection to search for specific classes, which is one of the \toolname's primary use cases. 
We should note though that not all of these reflection calls could be substituted with \toolname since there could be just not enough information at compile-time to do what is required (for example, when classes are needed to be created and used dynamically at run-time).

\Cref{tab:examples} highlights several popular (with at least 20 stars on GitHub) open-source Kotlin projects that can benefit from using \toolname. 
These projects come from different domains (graphical editors, static analyzers, test frameworks) and cover different contexts of the reflection usage, such as initializing the project or running tests.
The way reflection is applied also differs from project to project, including Java Reflection API itself as well as the \textit{reflections} and ClassGraph libraries. 
In these projects, all the cases of using reflection do not actually require run-time calculations, and therefore, they could be handled by \toolname.

\vspace{-0.5cm}
\subsection{Kotlin Compilation Process}\label{sec:compilation}

\subsubsection{Stages of the compilation process}

This subsection briefly summarizes key points of the Kotlin compilation process.

\begin{figure}[h]
    \centering
    \includegraphics[width=\columnwidth]{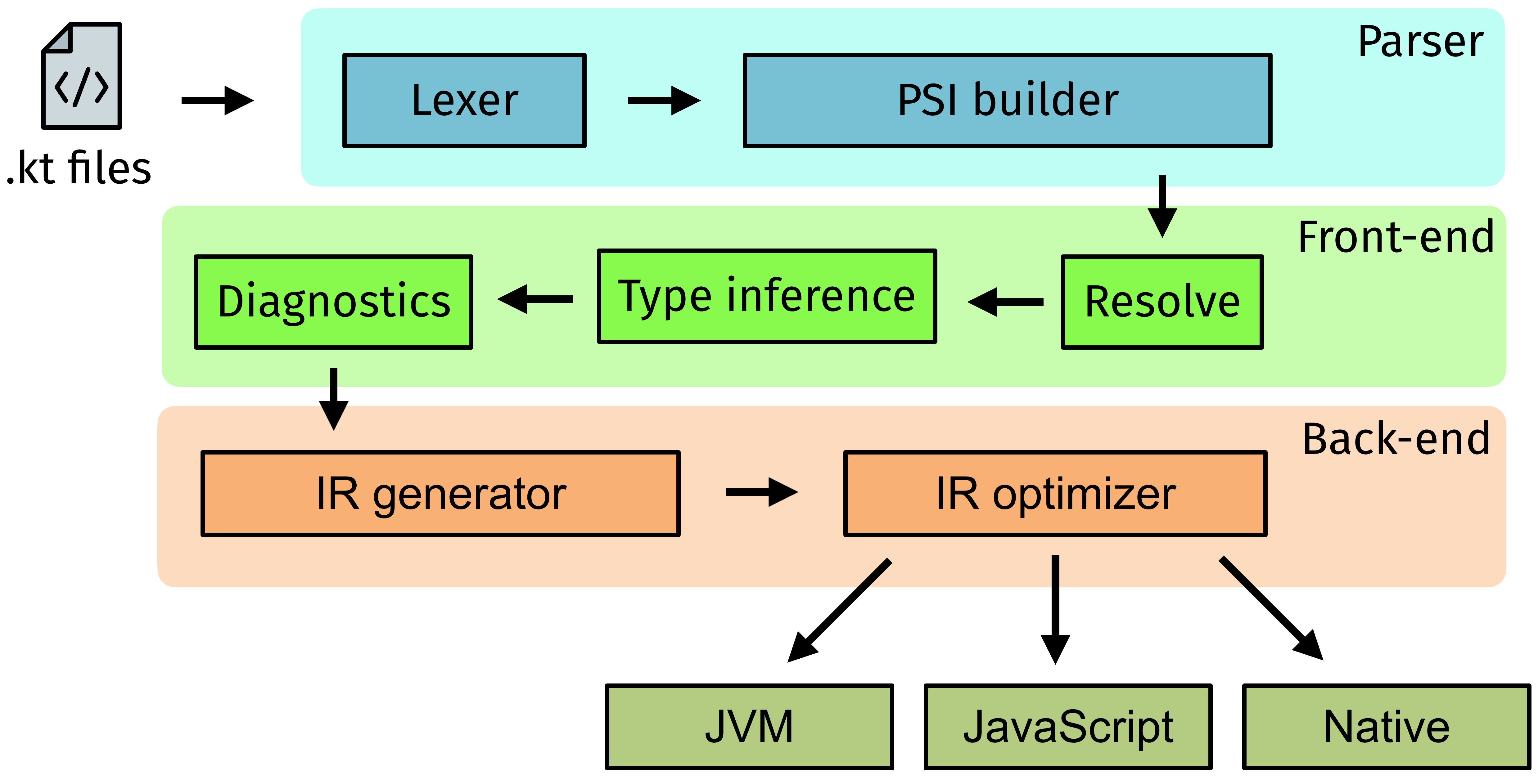}
    \caption{The main parts of the Kotlin compilation process.}
    \label{fig:compilation:process}
\end{figure}

% \smallskip
~\Cref{fig:compilation:process} presents the overall workflow of the Kotlin compilation process.
The Kotlin compiler has three main parts: parser, frontend and backend. 
The first part is responsible for parsing source files and building a rich syntax tree --- Program Structure Interface (PSI)~\cite{psi}. 
Next, the frontend part resolves the necessary dependencies, inferences all types, and runs compiler diagnostics to verify the correctness of the compiled code (\eg checks that the code does not contain two functions with the same signature and name). 
The backend part compiles the source code into a special intermediate representation (IR) and optimizes it.
Next, a platform-specific generator (JVM, JavaScript, or Kotlin Native) finishes the compilation process, \eg generates JVM bytecode.

\begin{figure*}[h]
    \centering
    \includegraphics[width=0.8\textwidth]{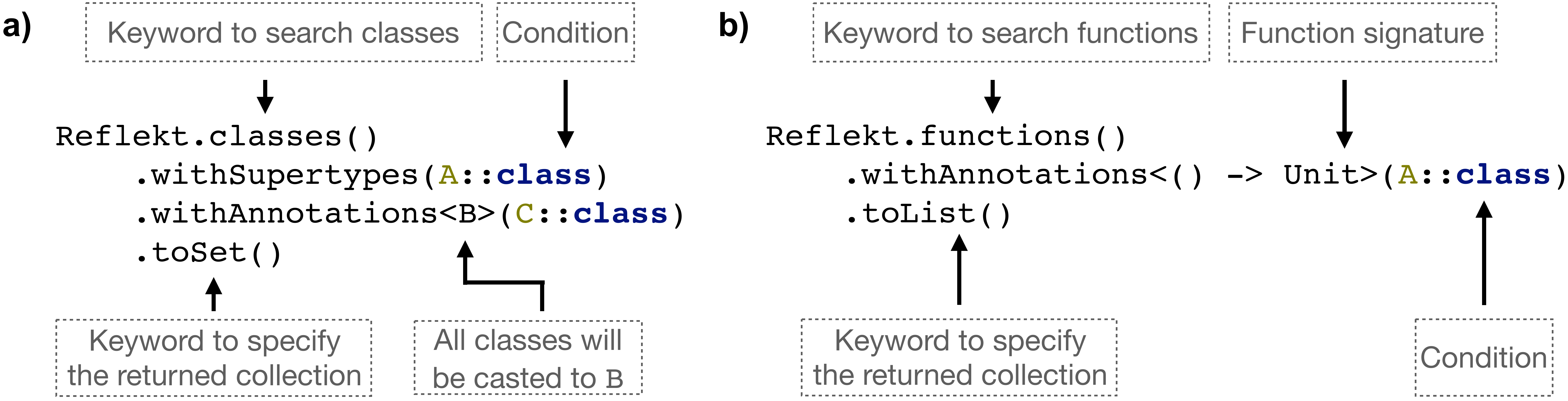}
    \caption{Examples of \toolname's core DSL: 
    \\ \textbf{a)} a \toolname query to find all classes with supertype \texttt{A}, annotation \texttt{C}, cast all of them to \texttt{B}, and put the result into a set. 
    \\ \textbf{b)} a \toolname query to find all functions without arguments, return type \texttt{Unit}, and annotation \texttt{A}. The result is returned as a list.}
    \label{fig:design:dsl:base}
\end{figure*}

\begin{figure*}[h]
    \centering
    \includegraphics[width=0.8\textwidth]{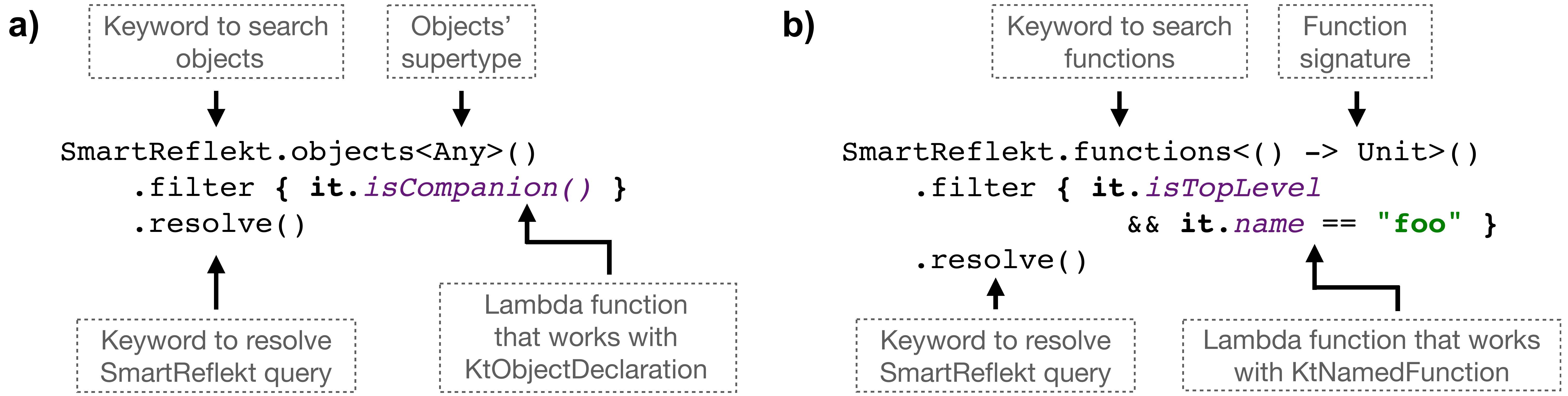}
    \caption{Examples of \toolname's extended DSL: 
    \\ \textbf{a)} a \smarttoolname query to find all companion objects with supertype \texttt{Any}. 
    \\ \textbf{b)} a \smarttoolname query to find all top-level functions with the name \texttt{foo}, return type \texttt{Unit}, and without any arguments.}
    \label{fig:design:dsl:extended}
\end{figure*}

\subsubsection{Kotlin compiler plugins}
The Kotlin compiler allows to write plugins, thus extending the internal functionality for all the supported platforms at any compilation stage. 
It has an incredibly powerful API with the access to all compilation stages, including intermediate representation of code, and the ability to modify internal classes and functions. 
For example, the \textit{kotlin-serialization} library~\cite{serialization} is implemented as a Kotlin compiler plugin to serialize classes.

\section{\toolname}\label{sec:design}

\toolname is a reflection library that provides the means to search for classes, object expressions, and functions in Kotlin source code at compile-time. 
It is written in Kotlin and implemented as a Kotlin compiler plugin to integrate itself into the compilation process.

\toolname consists of three main parts:
\begin{itemize}
    \item a domain-specific language (DSL) supporting two usage scenarios; 
    \item a Gradle plugin that facilitates the import of \toolname in end-users' projects;
    \item a Kotlin compiler plugin that implements compile-time resolution of reflection queries. 
\end{itemize}

Let us further discuss all these components in more detail.

\subsection{DSL}
One of the \toolname's main features is a concise and convenient way of its usage, which is not typical for other existing compile-time approaches (see~\Cref{subsubsection:compile-time} for an example).
This is achieved by providing a DSL that developers can use to find classes, object expressions, and functions in just one line of code.
In this context, object expressions are classes that implement the Singleton~\cite{gamma1995design} pattern, and thus have only one instance.

Due to the varying use cases and implementation differences, \toolname separates search conditions and DSLs provided for them into two types: \textit{core} and \textit{extended}.

\subsubsection{Core DSL} The most basic features of \toolname are:
\begin{itemize}
    \item finding classes and object expressions with specific annotations and supertypes;
    \item finding functions with specific annotations and signatures.
\end{itemize}

Examples of the core DSL queries are illustrated in~\Cref{fig:design:dsl:base}.

To use this DSL in a project, a developer should address the \texttt{Reflekt} object calling one of three methods: \texttt{classes()}, \texttt{objects()}, or \texttt{functions()} depending on the search intent. Then a search condition should be formed specifying needed annotations, supertypes, or function signatures accordingly.
The call chain concludes with a \texttt{toList()} or \texttt{toSet()} cast, which forms the result.

To specify annotations and supertypes, \texttt{KClass} objects should be used as parameters.
The \texttt{KClass} type is Kotlin’s counterpart to Java’s \texttt{java.lang.Class} and is used to represent Kotlin classes.
It can be obtained from any class by calling \texttt{::class} (for example, \texttt{MyClass::class}).

Function signatures are defined using the \texttt{FunctionN<*>} type, which is a Kotlin way to refer to functions.
For example, a function that takes an \texttt{Int} as an argument and returns a \texttt{Unit} (which acts as Java's \texttt{void}), would have a \texttt{(Int) -> Unit} signature.

\subsubsection{Extended DSL}
To support more advanced search conditions, we offer an extended DSL. 
It expands the core DSL with the ability to:
\begin{itemize}
    \item find classes and object expressions with a custom filter, including properties of the collected entities;
    \item find functions with a custom filter, including function properties.
\end{itemize}

Examples of the extended DSL are presented in~\Cref{fig:design:dsl:extended}.

This DSL is available through referencing the \texttt{SmartReflekt} object, where, similarly to the core DSL, one of the three methods can be called: \texttt{classes()}, \texttt{objects()}, or \texttt{functions()}.
In the type parameters of these methods, a supertype for object expressions or classes or a signature for functions can be specified.
Then there should be a lambda function (an analogue of anonymous functions in Java) defining a custom search condition, for instance,
being a companion for objects (Kotlin's counterpart for Java static functions) or being top-level for functions.
Finally, a concluding \texttt{resolve()} call is needed to return the found entities.

The supertype and the signature parameters should be of the same \texttt{KClass} and \texttt{FunctionN<*>} types as in the core DSL. As for the subsequent filtering conditions, the \texttt{filter()} function accepts parameters of \texttt{KtClass}~\cite{KtClass} for classes, \texttt{KtObjectDeclaration}~\cite{KtObjectDeclaration} for objects, and \texttt{KtNamedFunction}~\cite{KtNamedFunction} for functions. 
These are the Kotlin compiler's built-in types, which have a variety of properties, holding practically all known information about these entities. 
All this information can be used in a filtering predicate with the only limitation of not capturing any external variables. 
This restriction is required to guarantee that all condition parameters are available at compile-time. 

Implementation details of the extended DSL and the limitations they impose are discussed in~\Cref{subsubsection:library}.

\begin{figure}[h]
    \centering
    \includegraphics[width=1\columnwidth]{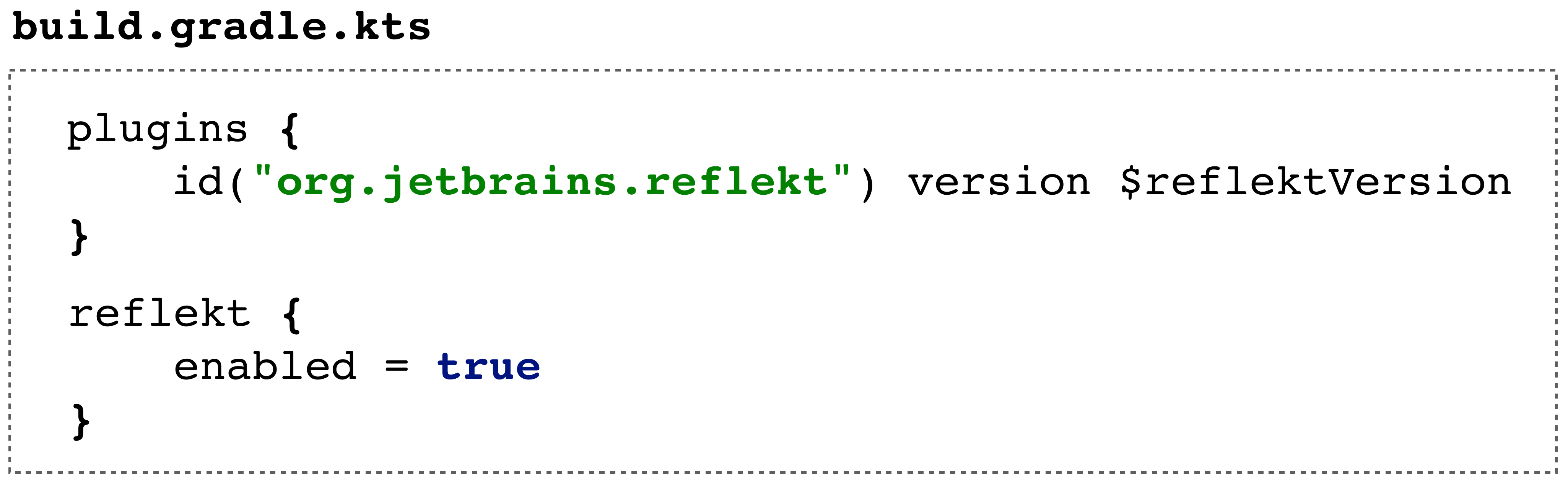}
    \caption{Setting up \toolname via Gradle in the \texttt{build.gradle.kts} file}
    \label{fig:gradle:setup}
\end{figure}

\subsection{Gradle Plugin}
Gradle is an open-source build automation tool that is designed to be flexible enough to build projects, in particular, written in Java and Kotlin~\cite{gradle}. 
\toolname as a library could be used quite easily using the Gradle plugin, which provides an entry point from a \texttt{build.gradle.kts} script and has a dependency on the module with \texttt{Reflekt} and \smarttoolname DSLs, allowing developers to use them in their projects. 
So, to use \toolname in a project, one simply needs to enable its Gradle plugin as shown in~\Cref{fig:gradle:setup}. 

\begin{figure}[b]
    \centering
    \includegraphics[width=0.7\columnwidth]{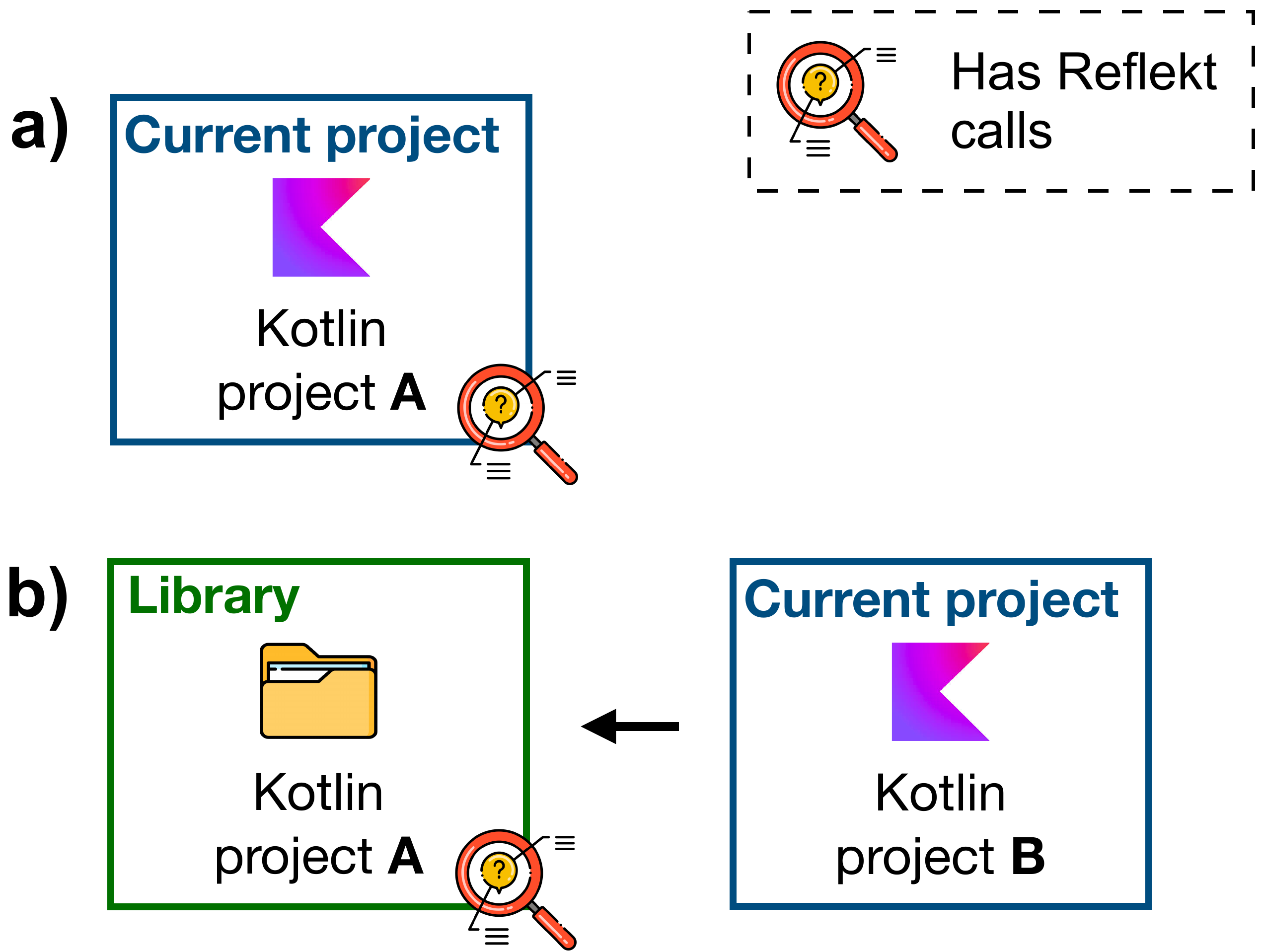}
    \caption{Two possible usage scenarios of \toolname: 
    \\ \textbf{a)} \toolname is used within a standalone project.
    \\ \textbf{b)} \toolname is used in a library \textbf{A}, which is used as a dependency in another project \textbf{B}.}
    \label{fig:design:piplines}
\end{figure}

\subsection{Compiler Plugin}\label{subsec:complier-plugin}
This component implements the main \toolname functionality and is designed as a Kotlin compiler plugin. 
Compiler plugins have an ability to intervene into the compilation process, including access to the intermediate representation (IR) of code: it is possible to analyze IR similarly to how abstract syntax trees are traversed and to even modify it, changing the code's final behavior.

There are two scenarios in which \toolname can be used (\Cref{fig:design:piplines}). 
The first one assumes a developer uses \toolname in some project to find entities within it. 
The second and more complicated one considers a case when the developed project is a library, and therefore the search should happen not only among the project itself, but also in other projects that add this library as a dependency. 
This section describes how these scenarios differ and what implementation decisions we made to support both use cases.

\begin{figure}[t]
    \centering
    \includegraphics[width=\columnwidth]{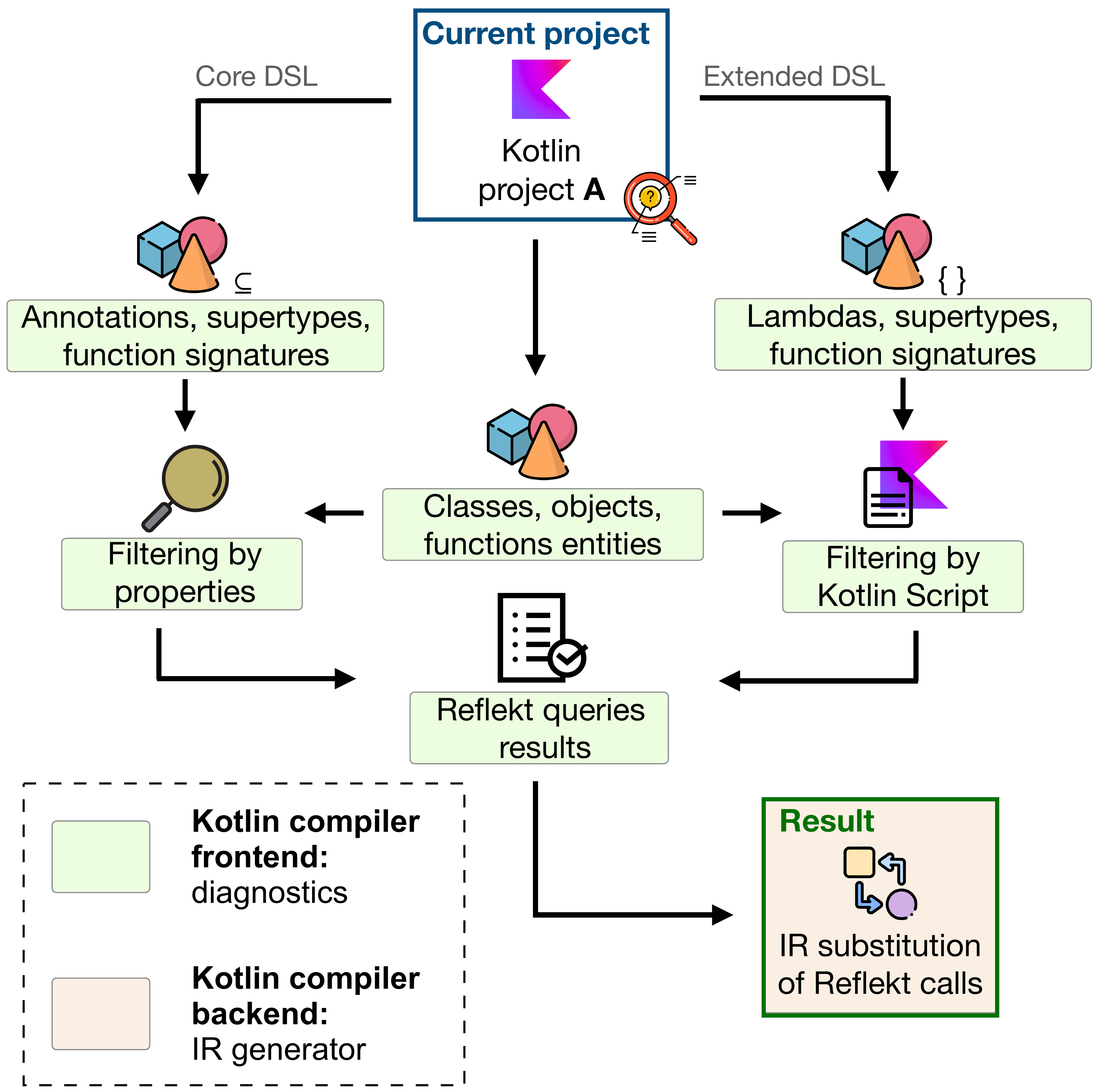}
    \caption{The general workflow of \toolname.}
    \label{fig:design:simple:pipline}
\end{figure}

\subsubsection{Using \toolname within a Standalone Project}\label{subsubsection:standalone}
This use case implies adding \toolname to the dependencies of some project to enable the search of entities within this project's source files.
In this scenario, both \toolnameAPI and \smarttoolname calls are supported.

The overall pipeline of \toolname is presented in~\Cref{fig:design:simple:pipline}.
The workflow starts with finding \toolnameAPI and \smarttoolname calls, then all of their parameters, such as types of instances to search, their supertypes, signatures, annotations, or custom lambda conditions are extracted. 
It happens on the \textit{diagnostics} stage of the compiler's frontend, since we need all qualified names already resolved for the subsequent search.

Next, all project's source files are scanned for the required entities that satisfy the search conditions.
One of the major limitations here comes from the nature of the Kotlin compilation process in multi-module projects: the compilation is performed for each module separately, so source files are not shared between modules during compilation. 
Therefore, in multi-module projects, \toolname can find entities only within each module.

The check for classes and object expressions' supertypes and function signatures is based on the Kotlin type system and subtype relations between them.  
For example, the \texttt{() -> Int} function signature satisfies the \texttt{() -> Any} search query, since the latter is more general than the former.
To do such type comparisons, Kotlin introduces the \texttt{KotlinType} class, 
full specification of the Kotlin types system can be found online~\cite{kotlinTypeSystem}. 

However, for other conditions that do not involve type checks, the matching is implemented differently for \toolnameAPI and \smarttoolname DSLs since the latter contains more complex filtering conditions.
If for the core DSL we can simply get all the information about entities from their internal representations (e.g., get all annotations for a specific \texttt{KtClass}), we cannot evaluate custom user condition in the same way since they are unknown in advance.
To support them, we employ the KotlinScript interpreter~\cite{kotlinScript}, which allows running lambda conditions from user queries at compile-time and filter out the entities that way.

\begin{figure}[t]
    \centering
    \includegraphics[width=\columnwidth]{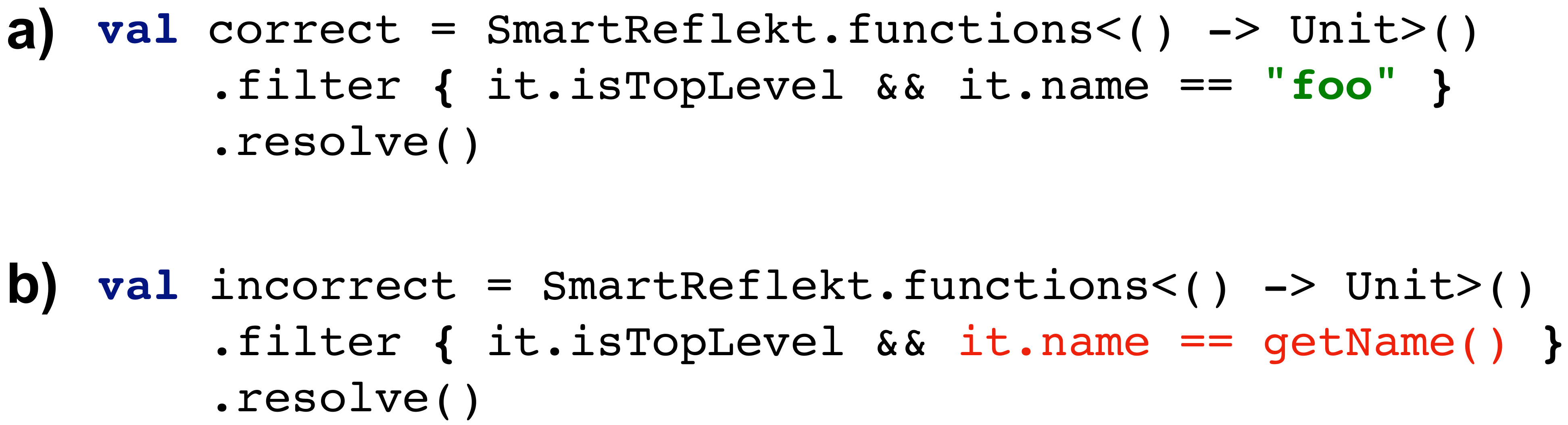}
    \caption{Example of a correct and incorrect \smarttoolname calls.\\
    \textbf{a)} a correct call without any external context inside lambda-conditions.\\
    \textbf{b)} an incorrect call if the \texttt{getName()} function comes from some external context.
    }
    \label{fig:lambda:examples}
\end{figure}

Since the KotlinScript interpreter is running during the compilation stage, no external variables should be used inside lambda conditions because they are unknown at this stage. 
However, external dependencies in the current project, which are available at compile-time and are included in Gradle using a \texttt{compileClasspath} configuration, are allowed in lambda conditions.
\Cref{fig:lambda:examples}a shows an example with the correct lambda body, which does not require any additional context. 
However, the filtering condition displayed in~\Cref{fig:lambda:examples}b is incorrect if the implementation of the \texttt{getName()} function is inaccessible to the KotlinScript interpreter. 

When all the instances that satisfy the search constraints are found, they should be returned from the \toolnameAPI and \smarttoolname calls.
To achieve that, the previous intermediate representation of these calls is substituted with the resulting list of the found entities.
It happens on the \textit{IR generation} compilation stage of the Kotlin compiler backend.

\subsubsection{Using \toolname within a Library}\label{subsubsection:library}
Another common use case assumes that \toolname is used within a library, so it is required to search for entities not only within the current project, but also among the entities of other projects that will later use this library. 
That is the exact scenario \kotless falls under since it is developed as a library that uses \toolname to search for entities in the projects of \kotless' end-users.

This scenario is fundamentally different from the previous one, since libraries are often developed and used by different people, and when a library is being compiled, there is no information about the code that will use it some time in the future. 
Therefore, there is a need to process \toolname calls during the compilation of the target project that uses the library, not the compilation of the library itself.
Another problem comes from the inability to substitute the library's IR, therefore, another way of changing the program's behavior should be implemented. 

\begin{figure}[b]
    \centering
    \includegraphics[width=0.8\columnwidth]{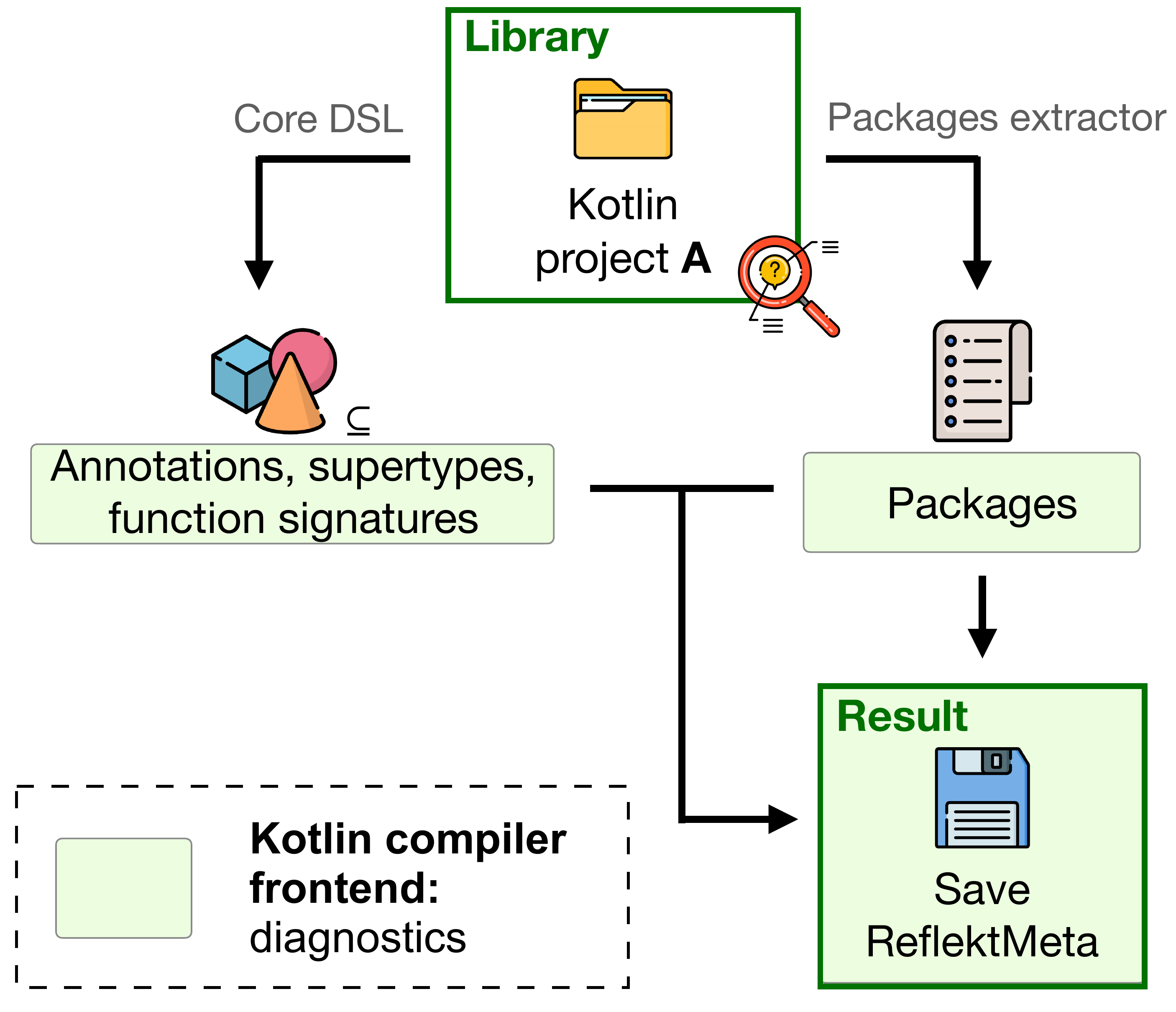}
    \caption{The pipeline of \toolname for the projects that will be used as libraries.}
    \label{fig:pipline:hard:library}
\end{figure}

In this scenario, we provide support only for the core DSL with \toolnameAPI calls due to the overall complexity of handling \smarttoolname calls. 
We plan to support the extended DSL for this scenario as well in the future. 

The workflow of \toolname in such cases is presented in~\Cref{fig:pipline:hard:library}. 
First, similar to the previous scenario, we start by gathering all \toolnameAPI calls when compiling the library.
But now these calls should be saved in order to be restored and used properly during the future project compilation to search among this project's entities. 
We save all the information about these calls in a separate \texttt{ReflektMeta} file.
The fact that such a file should be created for a given library is indicated in a Gradle configuration  file for this library.

The workflow continues when some project starts to use the developed library containing \toolnameAPI calls.
During the compilation of this project, the \texttt{ReflektMeta} file should be found and used. So to prevent from scanning all the libraries, those that are using \toolnameAPI calls should be explicitly listed in the project's Gradle configuration file.
Search conditions are then restored from the \texttt{ReflektMeta} file along with fully qualified names of the calls' parameters. 
The Kotlin compiler provides the means to get type descriptors from such names and then build appropriate \texttt{KotlinType} objects.
After that, the search for the entities within the current project is performed similarly to the previous scenario.

\begin{figure}[h]
    \centering
    \includegraphics[width=1\columnwidth]{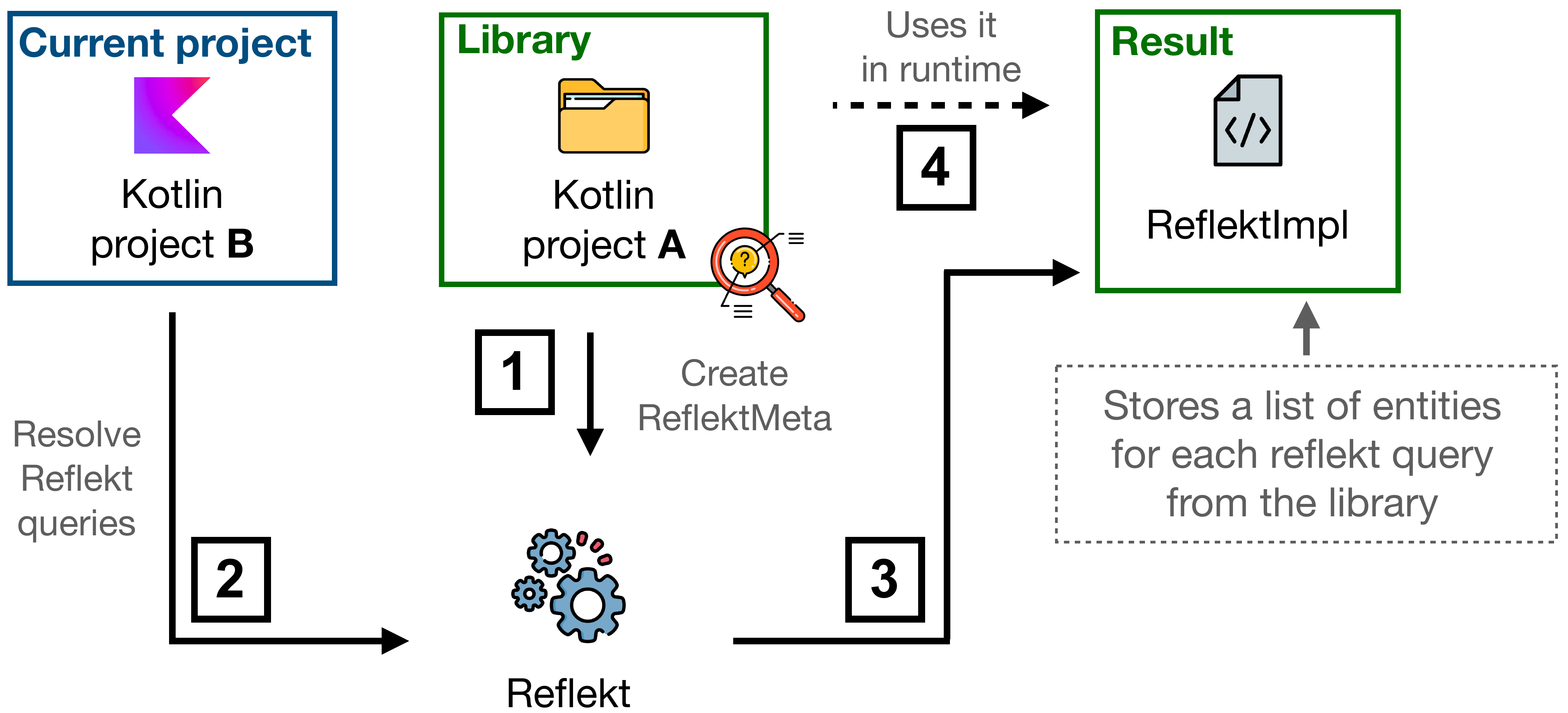}
    \caption{Using the \texttt{ReflektImpl} file to store the results of the search in the case when \toolname is used within a library.}
    \label{fig:pipline:reflektimpl}
\end{figure}

Another difficulty is that besides the current project, the search for the entities should be performed within the library as well, and at this point for this library there could be only compiled binaries present without any source code.
However, this problem can be efficiently solved by compiling a list of the existing packages in the library, which is also stored in the \texttt{ReflektMeta} file, utilizing the Kotlin compiler's ability to get all type descriptors from a given package.
The received descriptors then turn into \texttt{KotlinType}s, and the search continues in its original way among the library entities.

The last difference in the workflows of the two scenarios is the final step of substituting the initial behavior of \toolnameAPI calls to the found list of entities (see~\Cref{fig:pipline:reflektimpl}).
\toolname DSL is designed in a way that definitions of \toolnameAPI methods, which should return a list of the found entities, are stored in a separate file called \texttt{ReflektImpl.kt}.
By default, all these methods return an empty list.
It does not affect the first scenario (\Cref{subsubsection:standalone}), because there all \toolnameAPI calls are substituted with the actual found entities via changing the intermediate representation of code during compilation. 
However, there is no intermediate representation of code in the already compiled library. 
Nevertheless, we can substitute the whole \texttt{ReflektImpl.kt} file in the library with the new one, containing an appropriate implementation of methods, once this library gets used by some project. 
First, the original \texttt{ReflektImpl.kt} is excluded from the \toolname's \texttt{.jar} file. 
Then, during the compilation of the current project, once the result lists for every \toolnameAPI call are collected, a new \texttt{ReflektImpl.kt} is generated.
It provides an appropriate implementation for every \toolnameAPI call, returning the list of actual entities that exist in the source code. 
By including it back into the \toolname's \texttt{.jar} file, we make the library use the newly generated implementations instead of original reflection calls.

\section{Evaluation}\label{sec:evaluation}

We evaluated \toolname on the \kotless framework~\cite{tankov2021infrastructure} which is intended to simplify the development and deployment of serverless cloud applications written in Kotlin. 
The task of searching for objects and functions by specific conditions serves as a fundamental way to support code generation in this framework: \kotless scans user's project files, filters found entities by supertypes and annotations, and generates the necessary code for serverless deployment.
Moreover, \kotless is designed to be used as a library, so using \toolname in it would follow a scenario depicted in~\Cref{fig:pipline:hard:library}.

\begin{figure*}[!h]
        \centering
    \includegraphics[width=1\textwidth]{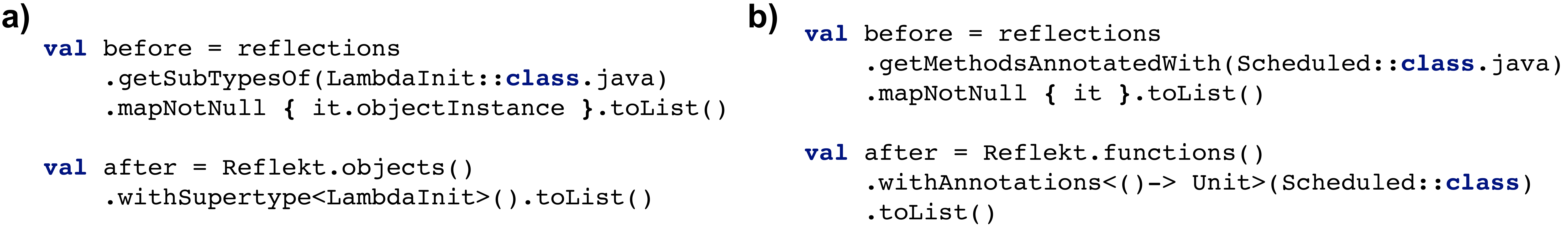}
    \caption{Examples of two \kotless reflection calls and their replacement with \toolname:  \\ a) Get all objects with a particular supertype. \\ b) Get all methods with a specific annotation and signature.}
    \label{fig:evaluation}
\end{figure*}

\begin{table*}[!h]
\centering
\begin{tabular}{lcccccccc}
    \toprule
    \multicolumn{1}{c}{\multirow{5}{*}{\textbf{Project}}} & \multicolumn{4}{c}{\textbf{Compilation (ms)}}  & \multicolumn{4}{c}{\textbf{Application start (ms)}} \\ \cmidrule(lr){2-5} \cmidrule(lr){6-9}
    % empty row
    % & & & & & & & & \\ 
    % compile\runtime
    & 
    \multicolumn{2}{c}{\textbf{Old approach}} & \multicolumn{2}{c}{\textbf{The Reflekt approach}}     & \multicolumn{2}{c}{\textbf{Old approach}} & \multicolumn{2}{c}{\textbf{The Reflekt approach}}     \\ \cmidrule(lr){2-3} \cmidrule(lr){4-5} \cmidrule(lr){6-7} \cmidrule(lr){8-9}
    % empty row
    % all\reflections
    & \multicolumn{1}{c}{\textbf{All}} & \multicolumn{1}{c}{\textbf{Reflection}} & \multicolumn{1}{c}{\textbf{All}} & \multicolumn{1}{c}{\textbf{Reflection}} & \multicolumn{1}{c}{\textbf{All}} & \multicolumn{1}{c}{\textbf{Reflection}} & \multicolumn{1}{c}{\textbf{All}} & \multicolumn{1}{c}{\textbf{Reflection}} \\ \midrule
    % content
    % Kotless
    % \multicolumn{1}{l}{\textbf{Kotless}} & 16,294 & - & - & - & 16,369 & \textbf{190 (1.2\%)} & - & - \\
     \multicolumn{1}{l}{\textbf{Kotless}} & $16,294 \pm 133$ & - & $16,369 \pm 120$ & $190 \pm 8$ (1.2\%) & - & - & - & - \\
    \multicolumn{1}{l}{}& & & & & & & & \\ [-0.8em]
    % site
    % \multicolumn{1}{l}{\textbf{Kotless site}} & 15,435 & - & 320 & 270 (84.4\%) & 15,590 & \textbf{173 (1.1\%)} & \textbf{265} & \textbf{220 (83\%)} \\
    \multicolumn{1}{l}{\textbf{Kotless website}} & $15,435 \pm 102$ & - &  $15,590 \pm 130$ & $173 \pm 11 $ (1.1\%) & $320 \pm 10$ & $270 \pm 10 (84.4\%)$ & $265 \pm 10$ & $220 \pm 10$ (83\%) \\
    \multicolumn{1}{l}{}& & & & & & & & \\ [-0.8em]
    % shorter
    % \multicolumn{1}{l}{\textbf{URL shorter}} & 16,194 & - & 290 & 235 (81\%) & 16,593 & \textbf{203 (1.2\%)} & \textbf{250} & \textbf{200 (80 \%)} \\
    \multicolumn{1}{l}{\textbf{URL shorter}} & $16,194 \pm 113$ & - & $16,593 \pm 160$ & $203 \pm 7$ (1.2\%) & $290 \pm 10$ & $235 \pm 10$ (81\%) & $250 \pm 10$ & $200 \pm 10$ (80 \%) \\
     \bottomrule 
\end{tabular}

    \vspace*{1.2mm} 
     \caption{Comparison of compilation and application start time for old and \toolname approaches together with the percentage of how much time the reflection calls take. }
     \label{tab:evaluation}
     \vspace{-7mm}
\end{table*}

\subsection{Replacement of Reflection Calls}

Currently, \kotless uses the \textit{reflections} library~\cite{reflections} for reflection queries. 
While some of them are meant to be performed at run-time due to the requirement of dynamic loading, others can be moved to compile-time. We replaced the latter with \toolname calls to measure the performance improvement. In total, $4$ reflection calls were replaced, examples of two of them are shown in~\Cref{fig:evaluation}. 

As can be seen, such a replacement does not require much effort due to the similarity of DSLs of \toolname and the \textit{reflections} library.

\subsection{Evaluation Setup} 

The repository of \kotless~\cite{kotlesslink} provides a set of example projects.
Two of them use only \kotless DSL~\cite{tankov2019kotless}, while others require additional frameworks such as Ktor~\cite{ktor} to create web applications.
For our experiments, we used only these two projects to eliminate possible interference with other libraries and frameworks.
The first project represents a static website containing \kotless' documentation~\cite{kotlesSite}.
The second one is a web service for shortening web URLs~\cite{kotlesShort}. 
Both of them rely on searching for objects by supertypes to initialize their web pages and searching for functions by annotations to construct the pages' routes.

For each example project, we measured the time spent on the compilation process and the application start together with the percentage of how much time it takes to process all the reflection calls.
The same measurements were performed after the reflection calls were replaced with \toolname queries. 
Some of the reflection calls could not be replaced with \toolname calls and therefore they still remain in the implementation of \kotless, affecting the application's start-up time.
Additionally, since \toolname increases the compilation time, 
we also study how the compilation time of \kotless itself has changed.
For each setting, we repeated the experiment 10 times, cleaning all the caches in between runs. 

To perform our experiments, we used an off-the-shelf MacBook laptop with the following characteristics: 2,4 GHz 8-Core Intel Core i9 processor and 32 GB 2667 MHz DDR4 RAM.

\subsection{Results}
The evaluation results are presented in~\Cref{tab:evaluation}.
It contains the measurements for both example projects, \textit{Kotless website} and \textit{URL shortener}, performed before and after embedding \toolname. 

Initially, there are no reflection queries at compile-time. 
However, \toolname moves the aforementioned reflection queries from run-time to compile-time. Thus, the compilation time slightly increases by about 1\% for both applications, as well as for the compilation of \kotless itself.

The most interesting part is the differences in applications' start times. 
First of all, we observe that reflection queries take a significant percentage of the start-up time, spanning from 80\% to almost 85\%.
However, the overall time of the applications' start decreases from 320~ms to 265~ms on average for the first project, and from 290~ms to 250~ms on average for the second one, achieving an improvement of 13.8\% and 17\% respectively. 
This performance boost is happening since the \textit{reflections} library takes considerable time to perform the classpath scanning for functions and objects. \toolname does not perform any scanning during run-time, so all applications are initialized faster, reducing cloud infrastructure costs for their end-users.

Moreover, the execution time of the \textit{reflections} library directly correlates with the number of classes it scans.
So, the larger the application is, the more improvement we can get when migrating from standard run-time reflection libraries to the proposed compile-time \toolname approach.

\section{Conclusion and future work}\label{sec:conclusion}

In this paper, we present \toolname: a plugin for the Kotlin compiler for compile-time reflection.
It allows searching for classes, object expressions, and functions by a given search condition at compile-time, overcoming the drawbacks of existing approaches.
It combines the efficiency of compile-time approaches with the convenient and concise DSL that these approaches lack.
At the same time, \toolname wins over the existing run-time approaches, which despite providing a convenient DSL, suffer from the poor performance. 
\toolname is implemented as a Kotlin compiler plugin which allows reusing the compiler internals to provide a compile-time reflection approach.

We evaluated the \toolname's performance on the \kotless framework, which has previously used the run-time \textit{reflections} library.
The measurements were performed on two serverless applications built using \kotless.
The results show an improvement in the applications' start time of 13.8\% and 17\% respectively.

The future work on \toolname will be aimed at overcoming its existing limitations.
First, we plan to support the search for entities in multiple modules at once, which is currently prohibited due to the nature of the Kotlin compilation process. 
We plan to achieve this by extracting descriptors of entities from the whole project and performing the search among them. 
In addition, we plan to optimize the analysis stage of the plugin to make it work even faster by improving the scanning mechanism and caching intermediate results.
We also intend to support incremental compilation~\cite{incrementalcompilation} once the required compiler API becomes available.
Another feature we want to implement is searching for entities among Java files, expanding \toolname usage to Java libraries and to multi-language projects.
Other plans include collecting the feedback from the end-users and implementing the requested features.
We already have several submitted requests asking for the ability to search for the sealed classes along with the other entities and to support \smarttoolname calls inside the libraries' files.

\section*{Acknowledgements}\label{sec:acknowledgements}

We thank Dmitriy Novozhilov and the rest of the Kotlin compiler team at JetBrains for their invaluable help and collaborative efforts.

\bibliographystyle{ACM-Reference-Format}
\balance
\bibliography{paper}

\end{document}